\documentclass[12pt]{article}
\usepackage[]{graphicx}
\usepackage[]{graphicx}
\usepackage{amsmath}
\usepackage{amssymb}
\usepackage{epsfig}
\usepackage{color}
\usepackage{subfigure}
\usepackage{lscape}
\usepackage{fullpage}
\usepackage{cite}

\def\##1{\underline{#1}}
\def\=#1{\underline{\underline{#1}}}

\def\r#1{(\ref{#1})}

\def\c#1{\cite{#1}}

\def\les{\left[}
\def\ris{\right]}
\def\lec{\left\{}
\def\ric{\right\}}

\def\.{\mbox{ \tiny{$^\bullet$} }}

\def\eps{\varepsilon}

\def\epso{\eps_{\scriptscriptstyle 0}}
\def\muo{\mu_{\scriptscriptstyle 0}}
\def\nuo{\nu_{\scriptscriptstyle 0}}
\def\etao{\eta_{\scriptscriptstyle 0}}
\def\tauo{\tau_{\scriptscriptstyle 0}}
\def\ko{k_{\scriptscriptstyle 0}}

\def\ro{(\#r,\omega)}
\def\rso{(\#r_s,\omega)}

\def\Vin{{\cal V}_{in}}
\def\Vout{{\cal V}_{out}}
\def\Surf{{\cal S}}
\def\OUT{_{out}}
\def\IN{_{in}}

\def\un{\hat{\#n}(\#r_s)}

 \def\ux{\hat{\#{u}}_x}
\def\uy{\hat{\#{u}}_y}
\def\uz{\hat{\#{u}}_z}

\def\oalpha{{\overline{\alpha}}}
\def\obeta{{\overline{\beta}}}
\def\oeps{{\overline{\eps}}}
\def\onu{{\overline{\nu}}}

\def\nr{\sqrt{(\nuo/\nu)}\sqrt{(\eps/\epso)}}
\def\deltar{\delta_r}
\def\etar{\eta_r}

\def\sp{\uy}
\def\pinc{\left(\frac{-\ux{\tauo}+\uz{\kappa}}{\ko}\right)}
\def\pref{\left(\frac{\ux{\tauo}+\uz{\kappa}}{\ko}\right)}
\def\ptr{\left(\frac{-\ux{\tau}+\uz{\kappa}}{k}\right)}

\def\as{a_s}
\def\ap{a_p}
\def\rs{r_s}
\def\rp{r_p}
\def\ts{t_s}
\def\tp{t_p}

\begin{document}

\begin{center}

\textbf{Axions, Surface States, and the Post Constraint in  Electromagnetics}\\

Akhlesh Lakhtakia$^1$ and Tom G. Mackay$^{1,2}$\\

\textit{$^1$Pennsylvania State University, Department of Engineering Science and Mechanics, \\
University Park, PA 16802, USA\\
$^2$University of Edinburgh, School of Mathematics, \\
Edinburgh EH9 3FD, Scotland, United Kingdom\\}

\end{center}

\noindent {\bf Abstract.} After formulating the frequency-domain Maxwell equations for a  homogeneous, linear, bianisotropic material occupying a bounded region, we found that the axionic piece vanishes from both the differential equations valid in the region and the boundary conditions, thereby vindicating the Post constraint. Our analysis indicates that characteristic effects that may be observed experimentally with magnetoelectric materials are not the consequences of the axionic piece but of an admittance that describes surface states.


\section{Introduction}
Do axions \cite{Raffelt} exist? We, the authors of this communication, lack the capabilities to answer that question. But we do note several negative answers to that question have emerged from experiments \cite{Zavattini,Robilliard,Asztalos,Roncadelli}. Does the ``axionic piece" \cite{OHpla}
of the linear constitutive dyadic exist in modern
classical electromagnetic theory? Again, we are not in a position to answer this question. But one of us summarized the developments before 2004 to state  that the \textit{recognizable} existence of the axionic piece is ruled out by modern
classical electromagnetic theory \cite{Lakh0}. The filtering out of the axionic piece (denoted by $\Psi$ in this
communication) by the Maxwell equations is enshrined
as the Post constraint $\Psi\equiv0$ \cite{LW1994,LW1996}, having been delivered by Post \c{Post} more than half a century ago.

The validity of the Post constraint has been  vigorously debated. A simple boundary-value problem has been formulated to show that $\Psi$ is measurable \cite{OHpla}. Multipole symmetry arguments have been derived to rule
against  as well as for the Post constraint \c{Raab1997,Raab2001,Raab2015}.
Data from scattering experiments conducted on a single ``Tellegen particle" were produced to show that $\Psi$ exists \cite{Tretyakov}, but those data---while sufficient to establish the magnetoelectric
phenomenon (which is not in doubt)---are insufficient to invalidate the Post constraint. Although suspensions of Tellegen particles have also been shown to exhibit the magnetoelectric phenomenon \cite{Ghosh1,Ghosh2}, no data has been presented to invalidate the Post constraint.

There is one item of experimental evidence against the Post constraint, as recounted in great detail by Hehl {\it et al.} \cite{HORSpra} in 2007. This evidence emerging from   {(i) certain assumptions of  symmetry of the linear
constitutive dyadic  that lead to the description of the magnetoelectric properties of  Cr$_2$O$_3$ in terms of just two scalars; (ii) DC measurements reported in 1994 to obtain the sign and the magnitude of one of the two magnetoelectric scalars as well as the magnitude of other magnetoelectric scalar} of Cr$_2$O$_3$  on a disk-shaped sample and a rectangular-solid sample \c{WJWRS1994}, respectively; (iii) the sign of the second magnetoelectric scalar of Cr$_2$O$_3$ measured at 10~kHz on a spherical sample reported in 1961 \c{Astrov1961}; and (iv) the experimentally supported approximation that   {the quasistatic permeability of Cr$_2$O$_3$  is the same
as that of vacuum} \c{Foner,HORSpra}, enabled them to deduce non-zero values of $\Psi$. Although independent confirmation of this experimental evidence does not exist to our knowledge, the veracity of measured data from careful experiments should not be doubted.

But the interpretation of measured data   {in order to obtain derivative quantities of greater significance} is not necessarily unimpeachable. It is our contention here that  data from electromagnetic experiments on samples of magnetoelectric materials
(such as Cr$_2$O$_3$) do not yield evidence  in favor of non-zero $\Psi$. Instead, the effects attributed to a non-zero $\Psi$, a macroscopic quantity that supposedly exists throughout a sample of a magnetoelectric material, arise from  surface states that exist   {due} to the abrupt  cessation of  microscopic morphology at the boundary of that sample.

The remainder of this communication is planned as follows: Section~\ref{vind} is devoted to a mathematical exposition of the Post constraint being a natural outcome of the application of Maxwell equations to linear materials. As $\Psi$ does not appear in the Maxwell equations, it is necessarily absent from the boundary conditions derived from those differential equations \c{Lakh2006}.   {Section~\ref{ss} presents the macroscopic characterization of surface states in terms of a surface charge density and a surface current density mediated by an admittance. Section~\ref{ie} contains the solution of a boundary-value problem to show that certain characteristic effects attributed to the axionic piece  are the consequences
of that admittance. This communication concludes with some remarks in Sec.~\ref{cr}. Vectors are underlined, 3$\times$3 dyadics \cite{EAB} are double underlined, $\#0$ is the null vector,  $\=I$ is the 3$\times$3 identity dyadic, and $\=0$ is the 3$\times$3 null dyadic, $\epso$ is the permittivity  of   vacuum,   and $\muo$ is the   permeability of vacuum.}

\section{Natural emergence of the Post constraint}\label{vind}
Suppose that all space is divided into two mutually disjoint regions $\Vout$ and $\Vin$ as well as the surface $\Surf$ that
separates $\Vout$ and $\Vin$. We formally distinguish between $\Vout$, $\Vin$, and $\Surf$. A differential equation equation holding in $\Vout$ (or $\Vin$) would have to be solved to satisfy boundary values of dependent
variables prescribed on $\Surf$.

Suppose that the region $\Vout$ is vacuous, and the region $\Vin$
is
occupied by a homogeneous, linear, bianisotropic material.
With an $\exp(-i\omega{t})$ dependence on time $t$
with angular frequency $\omega$ and $i=\sqrt{-1}$, the frequency-domain constitutive relations of free space are
specified as
\begin{equation}
\#D\ro=\epso\#E\ro\,,\qquad
\#H\ro=\nuo\#B\ro\,,\qquad \#r \in\Vout\,,
\label{conrel0}
\end{equation}
with $\nuo=\muo^{-1}$.  The frequency-domain constitutive relations
of the bianisotropic material are specified as
\begin{equation}
\left.\begin{array}{l}
\#D\ro=\=\eps(\omega)\.\#E\ro + \=\alpha(\omega)\.\#B\ro+\Psi(\omega)\#B\ro\\[4pt]
\#H\ro=\=\beta(\omega)\.\#E\ro + \=\nu(\omega)\.\#B\ro-\Psi(\omega)\#E\ro
\end{array}\ric\,,\qquad  {\#r \in\Vin}\,,
\label{conrel1}
\end{equation}
where
\begin{equation}
{\rm Trace}\les \=\alpha(\omega)- \=\beta(\omega)\ris=0\,.
\label{trace}
\end{equation}
The 3$\times$3  permittivity dyadic $\=\eps$, the 3$\times$3 impermeability dyadic $\=\nu$,  the two 3$\times$3 magnetoelectric dyadics $\=\alpha$
and $\=\beta$, and the axionic piece $\Psi$ can be combined into a single 6$\times$6  linear constitutive dyadic \cite{EAB},
but we do not use that compressed notation here.
When writing the constitutive relations \r{conrel1} for $\#D$ and $\#H$ in terms of $\#E$ and $\#B$, we have kept in mind that  $\#E$ and $\#B$ are primitive fields
because of their underlying microscopic existence while $\#D$ and $\#H$ are induction fields that arise at the macroscopic level but do not exist
at the microscopic level. The non-existence of induction fields at the microscopic level is  a cornerstone of
modern physics---in contrast to
a heterodox formulation of Hehl and Obukhov \c{HObook}
wherein the merely convenient but inessential $\#D$ and $\#H$ at the macroscopic level have microscopic counterparts,
thereby obscuring the distinction between the microscopic and
macroscopic levels.

Let us now apply the frequency-domain macroscopic Maxwell equations
\begin{equation}
\left.\begin{array}{l}
\nabla\.\#B\ro=0\\[4pt]
\nabla\times\#E\ro-i\omega\#B\ro=\#0\\[4pt]
\nabla\.\#D\ro=\rho\ro\\[4pt]
\nabla\times\#H\ro+i\omega\#D\ro=\#J\ro
\end{array}\ric\,
\label{ME}
\end{equation}
to both $\Vout$ and $\Vin$. After substituting Eqs.~\r{conrel0} in Eqs.~\r{ME}
we get
\begin{equation}
\left.\begin{array}{l}
\nabla\.\#B\ro=0\\[4pt]
\nabla\times\#E\ro-i\omega\#B\ro=\#0\\[4pt]
\epso\nabla\.\#E\ro=\rho\ro\\[4pt]
\nuo\nabla\times\#B\ro
+i\omega\epso\#E\ro=\#J\ro
\end{array}\ric\,,\quad{\#r\in\Vout}\,.
\label{MEout}
\end{equation}
Likewise, after substituting Eqs.~\r{conrel1} in Eqs.~\r{ME}
we get
\begin{equation}
\left.\begin{array}{l}
\nabla\.\#B\ro=0\\[4pt]
\nabla\times\#E\ro-i\omega\#B\ro=\#0\\[4pt]
\nabla\.\les\=\eps(\omega)\.\#E\ro + \=\alpha(\omega)\.\#B\ro+\Psi(\omega)\#B\ro\ris=\rho\ro\\[4pt]
\nabla\times\les\=\beta(\omega)\.\#E\ro + \=\nu(\omega)\.\#B\ro-\Psi(\omega)\#E\ro\ris\\[4pt]
\qquad\qquad+i\omega\les\=\eps(\omega)\.\#E\ro + \=\alpha(\omega)\.\#B\ro+\Psi(\omega)\#B\ro\ris=\#J\ro
\end{array}\ric\,,\quad{\#r\in\Vin}\,,
\label{MEin1}
\end{equation}
which   {can be first rearranged as
\begin{equation}
\left.\begin{array}{l}
\nabla\.\#B\ro=0\\[4pt]
\nabla\times\#E\ro-i\omega\#B\ro=\#0\\[4pt]
\nabla\.\les\=\eps(\omega)\.\#E\ro + \=\alpha(\omega)\.\#B\ro\ris +\Psi(\omega)\nabla\.\#B\ro
=\rho\ro\\[4pt]
\nabla\times\les\=\beta(\omega)\.\#E\ro + \=\nu(\omega)\.\#B\ro\ris 
-\Psi(\omega)\les\nabla\times\#E\ro-i\omega \#B\ro\ris
\\[4pt]
\qquad\qquad+i\omega\les\=\eps(\omega)\.\#E\ro + \=\alpha(\omega)\.\#B\ro\ris=\#J\ro
\end{array}\ric\,,\quad{\#r\in\Vin}\,,
\end{equation}
and then
simplified} to
\begin{equation}
\left.\begin{array}{l}
\nabla\.\#B\ro=0\\[4pt]
\nabla\times\#E\ro-i\omega\#B\ro=\#0\\[4pt]
\nabla\.\les\=\eps(\omega)\.\#E\ro + \=\alpha(\omega)\.\#B\ro\ris=\rho\ro\\[4pt]
\nabla\times\les\=\beta(\omega)\.\#E\ro + \=\nu(\omega)\.\#B\ro\ris\\[4pt]
\qquad\qquad+i\omega\les\=\eps(\omega)\.\#E\ro + \=\alpha(\omega)\.\#B\ro\ris=\#J\ro
\end{array}\ric\,,\quad{\#r\in\Vin}\,.
\label{MEin2}
\end{equation}
Most importantly, $\Psi$ does not appear in the Maxwell equations applied to $\Vin$ when the   merely convenient  induction
fields have been translated into primitive fields.

When solving an electromagnetic boundary-value problem, it is common to use the following boundary conditions
derived from Eqs.~\r{ME}:
\begin{equation}
\left.\begin{array}{l}
\un\.\les\#B\OUT\rso-\#B\IN\rso\ris
=0\\[4pt]
\un\times\les\#E\OUT\rso-\#E\IN\rso\ris=\#0\\[4pt]
\un\.\les\#D\OUT\rso-\#D\IN\rso\ris
=\rho_s\rso\\[4pt]
\un\times\les\#H\OUT\rso-\#H\IN\rso\ris=\#J_s\rso
\end{array}\ric\,,\qquad{\#r_s}\in\Surf\,,
\label{bc}
\end{equation}
with the unit vector $\un$ at $\#r_s\in\Surf$ pointing into $\Vout$. The quantities $\rho_s$ and $\#J_s$ are the surface charge density
and the surface current density, respectively. The subscripts $\IN$ and $\OUT$ indicate that the fields are being evaluated
on the interior and the exterior sides of $\Surf$. The boundary conditions \r{bc} are derived by integrating   {Eqs.~\r{ME}}
over pillboxes and closed contours straddling   {$\Surf$, as appropriate.}
Substituting the constitutive relations \r{conrel0} and \r{conrel1} in
Eqs.~\r{bc} and enforcing the charge neutrality and current neutrality of $\Surf$, we get
as the boundary conditions
\begin{eqnarray}
&&\left.\begin{array}{l}
\un\.\les\#B\OUT\rso-\#B\IN\rso\ris
=0\\[4pt]
\un\times\les\#E\OUT\rso-\#E\IN\rso\ris=\#0\\[4pt]
\un\.\Big[\epso\#E\OUT\rso-\=\eps(\omega)\.\#E\IN\rso \\
\qquad- \=\alpha(\omega)\.\#B\IN\rso-\Psi(\omega)\#B\IN\rso\Big]
=0\\[4pt]
\un\times\Big[\nuo\#B\OUT\rso-\=\beta(\omega)\.\#E\IN\rso \\
\qquad- \=\nu(\omega)\.\#B\IN\rso+\Psi(\omega)\#E\IN\rso\Big]=\#0
\end{array}\ric\,,\qquad{\#r_s}\in\Surf\,,
\label{bc1}
\end{eqnarray}
which clearly contain $\Psi$. Thus, one could conclude that although $\Psi$ has vanished from the Maxwell equations
applied to $\Vin$, it still makes its presence felt through the boundary conditions \r{bc1}.

That conclusion would be erroneous \c{Lakh2006}. As stated elsewhere \c{Lakh1}, ``the \textit{actual} [emphasis in the original] issue is the correct physical formulation of
boundary-value problems."
The boundary conditions \r{bc} are not sacrosanct in electromagnetics, but   their emergence from the Maxwell equations is.
Using the same pillboxes and closed contours as in the derivation procedure for the previous paragraph, but using
\begin{equation}
\left.\begin{array}{l}
\nabla\.\#B\ro=0\\[4pt]
\nabla\times\#E\ro-i\omega\#B\ro=\#0\\[4pt]
\nabla\.\les\=\oeps\ro\.\#E\ro + \=\oalpha\ro\.\#B\ro\ris=\rho\ro\\[4pt]
\nabla\times\les\=\obeta\ro\.\#E\ro + \=\onu\ro\.\#B\ro\ris\\[4pt]
\qquad\qquad+i\omega\les\=\oeps\ro\.\#E\ro + \=\oalpha\ro\.\#B\ro\ris=\#J\ro
\end{array}\ric\,
\label{ME2}
\end{equation}
with
\begin{equation}
\left.\begin{array}{ll}
\=\oeps\ro=\lec\begin{array}{l}\epso\=I\, \\ \=\eps(\omega)\,\end{array}\right.&\qquad
\=\oalpha\ro=\lec\begin{array}{l}\=0\, \\ \=\alpha(\omega)\,\end{array}\right.
\\[10pt]
\=\obeta\ro=\lec\begin{array}{l}\=0\,\\ \=\beta(\omega)\,\end{array}\right.&\qquad
\=\onu\ro=\lec\begin{array}{l}\nuo\=I\,\\ \=\nu(\omega)\,\end{array}\right.
\end{array}\ric\,,\qquad
\#r\in\lec\begin{array}{l}\Vout\,\\ \Vin\, \end{array}\right.\,,
\end{equation}
instead of Eqs.~\r{ME} because $\Psi$ vanishes from Eqs.~\r{MEin1} to deliver Eqs.~\r{MEin2}, we get
\begin{equation}
\left.\begin{array}{l}
\un\.\les\#B\OUT\rso-\#B\IN\rso\ris
=0\\[4pt]
\un\times\les\#E\OUT\rso-\#E\IN\rso\ris=\#0\\[4pt]
\un\.\les\epso\#E\OUT\rso-\=\eps(\omega)\.\#E\IN\rso - \=\alpha(\omega)\.\#B\IN\rso\ris
=0\\[4pt]
\un\times\les\nuo\#B\OUT\rso-\=\beta(\omega)\.\#E\IN\rso - \=\nu(\omega)\.\#B\IN\rso\ris=\#0
\end{array}\ric\,,\qquad{\#r_s}\in\Surf\,,
\label{bc2}
\end{equation}
as the boundary conditions. These boundary conditions do not contain $\Psi$, in contrast to the argument
raised elsewhere \cite{ST2008} based on the applicability of the Maxwell equations on $\Surf$ with $\Psi$ multiplied by
the unit step function with zero value for $\#r\in\Vout$.

What is the reason for the differences between the boundary conditions \r{bc1} and \r{bc2}? Very simply,
the boundary conditions \r{bc1} were derived from Eqs.~\r{ME} without considering that the redundancies  possibly
 contained in constitutive equations could be filtered out by the Maxwell equations when applied separately to $\Vin$
 and $\Vout$. When the redundancy
 in the form of $\Psi$ was filtered out by the Maxwell equations on application to $\Vin$, we obtained the boundary conditions \r{bc2} \cite{Lakh2006}.
 Thus, $ \Psi$ does not have a recognizable existence in either $\Vin$ or $\Vout$ and it does not enter the boundary conditions. Ergo, the Post constraint
 \begin{equation}
 \Psi(\omega)\equiv0
 \end{equation}
emerges as a natural outcome of the application of Maxwell equations to linear materials.

Let us   {stress here} that the standard boundary conditions
in macroscopic electromagnetics were not changed---in order to exclude
$\Psi$. That constitutive parameter was absent in the applicable differential equations (i.e., the Maxwell equations
containing the primitive but not the induction fields) and therefore did
not appear in the boundary conditions derived from those differential equations \c{Lakh2006}. Any suggestion
that the foregoing procedure
``actually \textit{changes} [emphasis in the original] the physical laws, namely,
the Maxwell equations at the interface between the two
[homogeneous] media" \cite{OH2009} appears strange to us because any differential equation is held to be valid in a region
but not on the boundary of that region.

 {\it Remark~1:} If $\Vin$ were to be occupied by a   homogeneous, linear, biisotropic material, then
 $\=\eps=\eps\=I$, $\=\nu=\nu\=I$, $\=\alpha=\alpha\=I=\=\beta$, and $\Psi=0$. In other words,
 a biisotropic material must be Lorentz~reciprocal and nonreciprocal biisotropy does not have a recognizable
 existence in electromagnetics \cite{LW1994}.

{\it Remark~2:} If $\Vout$ were to be occupied by a   homogeneous, linear, bianisotropic material different from the
 one occupying $\Vin$, the process of deriving Eqs.~\r{MEin2} and
 \r{bc2} shows that the Post constraint would apply to that material too.

 {\it Remark~3:} The Post constraint applies at every frequency, positive, negative, or zero.  Therefore, it also applies in the
 time domain.

{\it Remark~4:} Even though the existence of $\Psi$ may be deduced from crystallographic symmetry \c{Raab1997,Raab2015} or other non-electromagnetic means, it    {would still} be filtered out by the Maxwell equations.

   {\it Remark~5:} The so-called perfect electromagnetic conductor, characterized by $\#D=\Psi\#B$ and $\#H=-\Psi\#E$
  \cite{LS-PEMC,SL-PEMC-Rev}, does not have a recognizable existence as an electromagnetic medium because it violates the Post constraint.

 \section{Surface states}\label{ss}
The Post constraint does not negate the various manifestations of the linear magnetoelectric phenomenon
 as enshrined in the constitutive dyadics $\=\alpha$ and $\=\beta$, but it does constrain the constitutive characterization of that phenomenon.
In contrast, as is clear from Eqs.~\r{ME2}--\r{bc2}, $\Psi$ does not appear in the Maxwell equations in the two regions as well as
in the boundary conditions, when the constitutive relations are written for the correct induction fields ($\#D$ and $\#H$)   in terms of the
correct primitive fields ($\#E$ and $\#B$).
So, we sought an  alternative mechanism  that
would preserve the Post constraint but
could be invoked to interpret   {certain characteristics of}  experimental measurements on magnetoelectric materials.

 As in Sec.~\ref{vind}, let Eqs.~\r{conrel0} hold in $\Vout$ while the constitutive relations
\begin{equation}
\left.\begin{array}{l}
\#D\ro=\=\eps(\omega)\.\#E\ro + \=\alpha(\omega)\.\#B\ro \\[4pt]
\#H\ro=\=\beta(\omega)\.\#E\ro + \=\nu(\omega)\.\#B\ro
\end{array}\ric\,,\qquad   {\#r \in\Vin}\,,
\label{conrel1new}
\end{equation}
along with Eqs.~\r{trace} hold in $\Vin$. As boundary conditions, let us use Eqs.~\r{bc2} without assuming that
$\Surf$ is charge neutral and current neutral; i.e.,
\begin{equation}
\left.\begin{array}{l}
\un\.\les\#B\OUT\rso-\#B\IN\rso\ris
=0\\[4pt]
\un\times\les\#E\OUT\rso-\#E\IN\rso\ris=\#0\\[4pt]
\un\.\les\epso\#E\OUT\rso-\=\eps(\omega)\.\#E\IN\rso - \=\alpha(\omega)\.\#B\IN\rso\ris
=\rho_s\rso\\[4pt]
\un\times\les\nuo\#B\OUT\rso-\=\beta(\omega)\.\#E\IN\rso - \=\nu(\omega)\.\#B\IN\rso\ris=\#J_s\rso
\end{array}\ric\,,\qquad{\#r_s}\in\Surf\,.
\label{bc1new}
\end{equation}

We posit that the bianisotropic material  possesses
\textit{surface states} due to the abrupt cessation of its microscopic morphology, and that these surface states can be described macroscopically as
\begin{equation}
\left.\begin{array}{l}
\rho_s\rso = \gamma(\omega)\,\un\.\#B\IN\rso\\[4pt]
\#J_s\rso=-\gamma(\omega)\,\un\times\#E\IN\rso
\end{array}\ric\,,\qquad{\#r_s}\in\Surf\,,
\label{scdef}
\end{equation}
where $\gamma$ is an admittance characteristic of the material. What may be observed experimentally with magnetoelectric materials are not the effects of
a constitutive parameter $\Psi$ that describes electromagnetic phenomenons in a volume, but of an admittance $\gamma$ that  describes electromagnetic phenomenons on a surface.

{\it Remark~6:} If $\Vout$ were to be occupied by a   homogeneous, linear, bianisotropic material different from the
 one occupying $\Vin$, and both materials were to possess surface states, then
 \begin{equation}
\left.\begin{array}{l}
\rho_s\rso =  \un\.\les\gamma\IN(\omega)\#B\IN\rso-\gamma\OUT(\omega)\#B\OUT\rso\ris\\[4pt]
\#J_s\rso=-\un\times\les\gamma\IN(\omega)\#E\IN\rso-\gamma\OUT(\omega)\#E\OUT\rso\ris
\end{array}\ric\,,\qquad{\#r_s}\in\Surf\,,
\label{scdef-a}
\end{equation}
because the unit normal vector at $\#r_s\in\Surf$ pointing into $\Vin$ is  $-\un$.

\section{Illustrative Example}\label{ie}
In order to illustrate the effects of $\gamma$, let us consider a simple boundary-value problem that has been put forward
to indicate the measurability of $\Psi$ \cite{OHpla}. Let us actually begin with a slightly more complicated problem \cite{ST2008} to be simplified later.

Suppose that  $\Vout$   is the half space $z<0$, $\Vin$   is the half space $z>0$, and therefore $\Surf$   is the plane
$z=0$. Furthermore, let $\=\eps=\eps\=I$, $\=\nu=\nu\=I$, $\=\alpha=\=0$, and $\=\beta=\=0$, where the dependency upon $\omega$ has been omitted for brevity.

Let  the primitive electromagnetic
fields in $\Vout$ be
 \begin{equation}
\left.\begin{array}{l}
\#E(\#r)=\Big\{ \les\as\,\sp +\ap \,\pinc\ris \exp(i\tauo{z})
\\[6pt]
\qquad\qquad+\les\rs\,\sp +\rp \,\pref\ris \exp(-i\tauo{z})
\Big\}
\exp(i\kappa {x})
\\[10pt]
\#B(\#r)=\frac{\ko}{\omega}\Big\{ \les-\ap\,\sp +\as \,\pinc\ris \exp(i\tauo{z})
\\[6pt]
\qquad\qquad+\les-\rp\,\sp +\rs \,\pref\ris \exp(-i\tauo{z})
\Big\}
\exp(i\kappa {x})
\end{array}\right\}
\, , \qquad z < 0
\, .
\label{EBincref}
\end{equation}
where $\ko=\omega\sqrt{\epso/\nuo}$ and $\tauo=+\sqrt{\ko^2-\kappa^2}$.
Representing a plane wave incident on $\Surf$,
the coefficients $\as$ and $\ap$ are presumed to be known. Representing the plane wave
reflected into $\Vout$, the coefficients $\rs$ and $\rp$ are
unknown. Equations~\r{EBincref} satisfy the homogeneous counterparts of Eqs.~\r{MEout}.

The primitive electromagnetic fields in $\Vin$ are given as
 \begin{equation}
\left.\begin{array}{l}
\#E(\#r)=  \les\ts\,\sp +\tp \,\ptr\ris \exp(i\tau{z})
\exp(i\kappa {x})
\\[10pt]
\#B(\#r)=\frac{k}{\omega} \les-\tp\,\sp +\ts \,\ptr\ris \exp(i\tau{z})
\exp(i\kappa {x})
\end{array}\right\}
\, , \qquad z > 0
\, ,
\label{EBtr}
\end{equation}
where $k=\ko\nr$, $\tau = + \sqrt{k^2 - \kappa^2}$,
and
 the coefficients $\ts$ and $\tp$ are unknown.
 Representing the plane wave refracted
into $\Vin$,
these expressions satisfy the homogeneous counterparts of Eqs.~\r{MEin1} and \r{MEin2}.

The foregoing expressions were substituted in Eqs.~\r{bc1new}$_{2,4}$ and \r{scdef}$_2$ to determine
$\rs$, $\rp$, $\ts$, and $\tp$ in terms of $\as$ and $\ap$. The results found are as follows:
\begin{eqnarray}
\rs&=& \as
\frac{(\etar-\deltar)(1+\etar\deltar)-(\gamma\etao)^2\etar^2\deltar}{(\etar+\deltar)(1+\etar\deltar)+(\gamma\etao)^2\etar^2\deltar}
+
\ap
\frac{2\gamma\etao\etar^2\deltar}{(\etar+\deltar)(1+\etar\deltar)+(\gamma\etao)^2\etar^2\deltar}\,,
\label{rs-def}
\\
\rp&=& \ap
\frac{(\etar+\deltar)(1-\etar\deltar)+(\gamma\etao)^2\etar^2\deltar}{(\etar+\deltar)(1+\etar\deltar)+(\gamma\etao)^2\etar^2\deltar}
+
\as
\frac{2\gamma\etao\etar^2\deltar}{(\etar+\deltar)(1+\etar\deltar)+(\gamma\etao)^2\etar^2\deltar}\,,
\label{rp-def}
\\
\ts&=&\as
\frac{2\etar(1+\etar\deltar)}{(\etar+\deltar)(1+\etar\deltar)+(\gamma\etao)^2\etar^2\deltar}
+
\ap
\frac{2\gamma\etao\etar^2\deltar}{(\etar+\deltar)(1+\etar\deltar)+(\gamma\etao)^2\etar^2\deltar}\,,
\label{ts-def}
\\
\tp&=& \ap
\frac{2\etar(\etar+\deltar)}{(\etar+\deltar)(1+\etar\deltar)+(\gamma\etao)^2\etar^2\deltar}
-
\as
\frac{2\gamma\etao\etar^2}{(\etar+\deltar)(1+\etar\deltar)+(\gamma\etao)^2\etar^2\deltar}\,,
\label{tp-def}
\end{eqnarray}
where
\begin{equation}
\deltar=\frac{\tau/k}{\tauo/\ko}\,,\qquad
\etar=\sqrt{\frac{\epso\nuo}{\eps\nu}}\,.
\end{equation}
We have verified that Eqs.~\r{rs-def}--\r{tp-def} satisfy Eqs.~\r{bc1new}$_{1,3}$ and \r{scdef}$_1$;
furthermore, these expresssions simplify to the standard results
\begin{equation}
\left.\begin{array}{ll}
\rs = \as
\frac{\etar-\deltar}{\etar+\deltar}
\,,\qquad
&\rp= \ap
\frac{1-\etar\deltar}{1+\etar\deltar}
\\[8pt]
\ts=\as
\frac{2\etar}{\etar+\deltar}
\,,\qquad
&\tp=\ap
\frac{2\etar}{1+\etar\deltar}
\end{array}\ric\,
\end{equation}
for $\gamma=0$ \c{Lakh2}.

Cross-polarized reflection in the foregoing
problem has been used to adduce the invalidity of the
Post constraint \cite{ST2008}, but   Eqs.~\r{rs-def}--\r{tp-def} show that both cross-polarized reflection and refraction are due instead to the surface states as accommodated
phenomenologically through $\gamma$.

Let us now simplify the results to apply to the  boundary-value problem   put forward
to indicate the measurability of $\Psi$ \cite{OHpla}. For that problem $\eps=\epso$ and $\nu=\nuo$.  Then,
$\deltar=\etar=1$ and Eqs.~\r{rs-def}--\r{tp-def} simplify to:
\begin{equation}
\left.\begin{array}{ll}
\rs= -\gamma\etao
\frac{\as\gamma\etao-\ap}{4+(\gamma\etao)^2}
\,,\qquad &
\rp= \gamma\etao
\frac{\ap\gamma\etao+\as}{4+(\gamma\etao)^2}
\\[8pt]
\ts=2
\frac{2\as+\ap\gamma\etao}{4+(\gamma\etao)^2}
\,,\qquad &
\tp=2
\frac{2\ap-\as\gamma\etao}{4+(\gamma\etao)^2}
\end{array}\ric\,.
\label{soln-OHpla-prob}
\end{equation}
As both half spaces $z<0$ and $z>0$ are vacuous, Eqs.~\r{soln-OHpla-prob} indicate that (i) reflection occurs
solely due to non-zero $\gamma$ and (ii) the polarization  of the reflected plane wave is rotated with respect to
that of the incident plane wave. Both characteristics had been attributed to a non-zero $\Psi$\cite{OHpla}, but that
is clearly unnecessary.

\section{Concluding remarks}\label{cr}
We formulated  the frequency-domain Maxwell equations for a  homogeneous, linear, bianisotropic material occupying a bounded region, as well as for the vacuous region that makes up the rest of space. The axionic piece $\Psi$ is absent in vacuum and it disappears from the Maxwell equations in the bianisotropic material when the induction fields $\#D$ and $\#H$ are replaced by the primitive fields $\#E$ and $\#B$ by using the constitutive equations. In consequence, the axionic piece also vanishes from the boundary conditions prevalent at the interface of the two regions. The Post constraint $\Psi\equiv0$ is thereby vindicated. 

Even if both regions were were to be occupied by dissimilar   homogeneous, linear, bianisotropic materials, the derivation process indicates that   the Post constraint would apply to both materials. Even though the existence of $\Psi$ may be deduced somehow, it will be filtered out by the Maxwell equations. Furthermore, every
biisotropic material must be Lorentz~reciprocal and the so-called perfect electromagnetic conductor does not have a recognizable existence as an electromagnetic medium because it violates the Post constraint.

We also found that characteristic effects that may be observed experimentally with slabs of a magnetoelectric material are not the consequences of the axionic piece of the linear constitutive dyadic of that material. Instead, they are the consequences
of an admittance that describes surface states arising from  the abrupt cessation of the microscopic morphology of the
magnetoelectric material. This admittance will also impact classical analyses \c{Chang,Liu} of light scattering by topological
insulators. \c{Hasan}

\end{document}